\documentclass{iauc}
\usepackage{graphicx}
\pubyear{2005}
\volume{199}
\pagerange{1--}
\jname{Probing Galaxies through Quasar Absorption Lines}
\editors{P. R. Williams, C. Shu, and B. M\'{e}nard, eds.}

\title[3-D matter distribution at $z\sim2$ ] 
{Probing 3-D matter distribution at $z\sim2$ with QSO multiple lines
  of sight\footnote{Based on observations collected at the European
  Southern Observatory, Chile}}

\author[S. Cristiani, V. D'Odorico, F. Saitta, M. Viel et al.]   
{S. Cristiani$^1$,
V. D'Odorico$^1$, F. Saitta$^2$, M. Viel$^3$, S. Bianchi$^4$, \break
B. Boyle$^5$, S. Lopez$^6$, J. Maza$^6$, P. Outram$^7$}

\affiliation{$^1$INAF - Osservatorio Astronomico di Trieste, via
  G.B. Tiepolo, 11 I-34131 Trieste, Italy, \break email: cristiani@ts.astro.it,
dodorico@ts.astro.it   
\\[\affilskip]
$^2$Dipartimento di Astronomia, Universit\`a degli Studi di Trieste, via
  G.B. Tiepolo, 11 \break I-34131 Trieste, Italy  
\\[\affilskip]
$^3$Institute of Astronomy, Madingley Road, Cambridge CB3 0HA 
\\[\affilskip]
$^4$INAF-Istituto di Radioastronomia/Sez. Firenze, Largo Enrico Fermi,
  5, \break I-50125 Firenze, Italy 
\\[\affilskip]
$^5$Australia Telescope National Facility, PO Box 76, Epping NSW 1710,
  Australia 
\\[\affilskip]
$^6$Departamento de Astronomia, Universidad de Chile, Casilla 36-D,
  Santiago, Chile 
\\[\affilskip]
$^7$Department of Physics, University of Durham, South Road, Durham,
  DH1 3LE 

}

\begin{document}

\maketitle

\begin{abstract}
We investigate the 3-D matter distribution at $z \sim 2$ with high
resolution ($R\sim 40000$) spectra of QSO pairs and groups obtained
with the UVES spectrograph at ESO VLT.  Our sample is unique for the
number density of objects and the variety of separations, between
$\sim 0.5$ and 7 proper Mpc\footnote{We adopt a $\Lambda$CDM cosmology
with $\Omega_m = 0.3$, $\Omega_{\Lambda}=0.7$ and $h=0.72$}.  We
compute the real space cross-correlation function of the
Lyman-$\alpha$ forest transmitted fluxes.
There is a significant clustering signal up to $\sim 2$ proper Mpc,
which is still present when absorption lines with high column density
($\log N \ge 13.8$) 
are excluded.

\end{abstract}

\firstsection 
\section{Introduction}

The Ly-$\alpha$ forest absorbers detected at $z > 1.5$ in high
resolution QSO spectra outnumber any other population of objects
observable from the ground at those redshifts.  They originate, as
shown by hydro-simulations (e.g. \cite[Cen \etal\ 1994]{cen1994}), in
density fluctuations of the intergalactic medium (IGM), i.e. in the
web of filamentary structures connecting the densest peaks, and are
reliable tracers of the baryon density field as well as of the
underlying dark matter distribution.

From the study of absorption spectra along single lines of sight (LOSs)
to distant QSOs it has been possible to determine the shape and
amplitude of the power spectrum of the spatial distribution of dark
matter (e.g., \cite[Kim \etal\ 2004]{kim04}, Viel, this conference) 
or gain important
information on the baryon density of the Universe (\cite[Rauch \etal\
1997]{rauch97}) and on the physical state of the
IGM (\cite[Schaye \etal\ 2000]{schaye2000}).

Our present analysis introduces the use of adjacent LOSs 
to probe the actual 3-D distribution of matter in the Universe 
and provide estimates of the size/correlation of the absorbing structures. 
Hydrodynamical (\cite[Charlton \etal\ 1997]{charlton97}) and analytical 
(\cite[Viel \etal\ 2002]{viel2002}) simulations have shown the advantages 
in using multiple LOSs with respect to single ones, in particular, 
to decrease the effects of inaccuracies in the continuum fitting. 
    
\section{Our sample of QSOs}

Two major breakthroughs have dramatically improved the exploitation of
the potential offered by QSO multiple LOSs:
the 2dF QSO Redshift Survey (2QZ, \cite[Croom \etal\ 2004]{croom04}),
whose complete spectroscopic catalog contains more than $\sim
23000$ QSOs, approximately 50 times more than the previous largest
QSO survey to a similar depth ($B<21$) and the UVES spectrograph at
the ESO VLT telescope which has a remarkable efficiency especially in
the extreme UV. 

We have  searched the 2dF and other QSO databases for the best groups 
with apparent
magnitude $B\le 20$ and $z>1.8$ and carried on a great observational 
effort to collect UVES spectra of the selected QSOs. 
At the moment, we have observed one QSO pair, one triplet and one
sextet of QSOs with a resolution $R\sim 40000$ and signal-to-noise
ratio in the Lyman-$\alpha$ forest larger than 3. 
With the addition of two more UVES QSO pairs from our archive, we
total 20 different baselines with proper spatial separations between 
$\sim 0.5$ and 7 Mpc, a unique sample both for the number
density and the variety of LOS separations investigated. 

\section{The transmitted flux cross-correlation function}

\begin{figure}
\includegraphics[height=9truecm,width=10truecm]{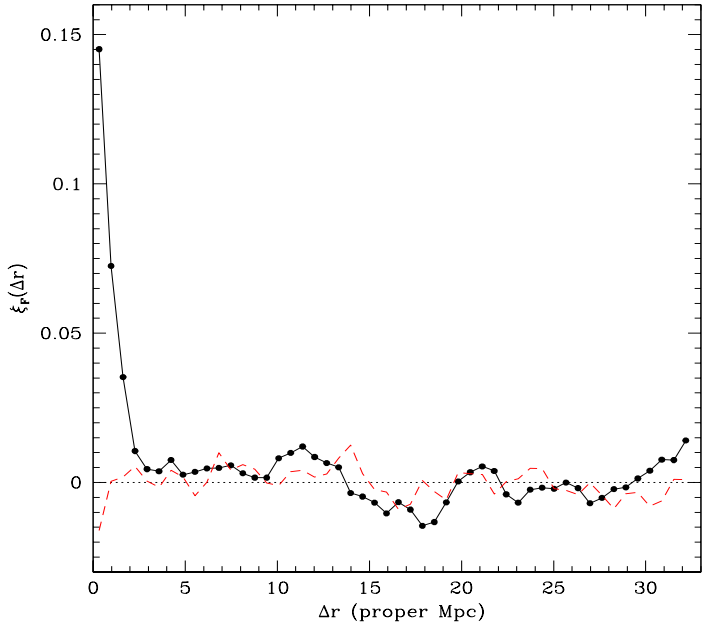}
  \caption{Correlation function of the transmitted flux in real space
    for our sample of pairs and groups of QSOs. The dashed line is the 
    result obtained for the uncorrelated control sample (see text)}\label{fig:spcf}
\end{figure}

We select in each normalized spectrum the region between the 
Lyman-$\beta$ emission (or the minimum observed wavelength, when the 
Lyman-$\beta$ falls outside the spectrum) and 5000 km s$^{-1}$ 
from the Lyman-$\alpha$ emission (to avoid proximity effect) 
and compute the cross-correlation function of the transmitted flux 
between adjacent LOSs. 
The following generalization of the formula for the 
density field (e.g. \cite[Peebles 1980]{peebles1980}) has been adopted:

\begin{equation}
\xi_F(\Delta r) = \frac{ < (F(x) - <F>)(F_0(x+\Delta r) - <F_0>)>}
{ \sqrt{< (F(x) - <F>)^2><(F_0(x+\Delta r) - <F_0>)^2>}}
\end{equation}

\noindent
where $F$ and $F_0$ are the transmitted fluxes in two adjacent LOSs 
and $\Delta r$ is the proper spatial separation between two pixels 
in different LOSs. The cross-correlation function is estimated 
over all the considered pixels of all the QSO pairs in the sample. 
To check for the effects of systematics we also computed the same
cross-correlation function on a control sample, obtained substituting
in turn one of the QSO spectra in each baseline with an 'uncorrelated'
one, roughly at the same redshift, but far away on the sky, picked up
in our database of high-resolution, high-S/N QSO spectra. 
The result is shown in Fig.~\ref{fig:spcf}. A correlation signal is 
present up to proper separations of $\sim 2$ Mpc indicating a large 
coherence length of the absorbers. 
This length is consistent with the size of Lyman-$\alpha$ absorbers 
we found in \cite[D'Odorico \etal\ (1998)]{vale98} with the line analysis 
of coincidences and anti-coincidences. 


In order to verify the relative contribution of the stronger and weaker 
absorption lines to the clustering signal we have selected the pixels with 
$F > 0.2$ corresponding to $\log N({\rm HI}) < 13.7-13.8$ and recomputed  
$\xi_F(\Delta r)$. Although significantly decreased (of a factor $\sim
2$) a clustering signal is still clearly present.  
This is an indication that also far from the most over-dense regions, 
in the 'true' IGM, matter is still clustered. 
   
A comparison with hydrodynamical simulations (see, for a description,
Viel et al. 2004) shows a substantial agreement between the predicted
and observed signal.



\begin{thebibliography}{}

\bibitem[]{charlton97} 
{Charlton, J.C., Anninos, P., Zhang, Y., Norman, M.L.} 1997, 
\textit{ApJ} 485, 26

\bibitem[Cen \etal\ (1994)]{cen1994}
{Cen, R., Miralda-Escud\'e, J., Ostriker, J.P., Rauch, M.} 1994, 
\textit{ApJ} 437, L83 


\bibitem[]{croom04}
{Croom, S. M., Smith, R. J., Boyle, B. J., Shanks, T., Miller, L.,
Outram, P. J., Loaring, N. S.} 2004, 
\textit{MNRAS} 349, 1397
 

\bibitem[D'Odorico \etal\ (1998)]{vale98}
{D'Odorico, V., Cristiani, S., D'Odorico, S., Fontana, A., Giallongo, E., 
Shaver, P.} 1998, 
\textit{A\&A} 339, 678

\bibitem[]{kim04}
{Kim, T.-S., Viel, M., Haehnelt, M., Carswell, R.F., Cristiani, S.}
  2004, 
\textit{MNRAS} 347, 355

\bibitem[]{peebles1980}
{Peebles, P.J.E.} 1980, 
in \textit{The Large-Scale Structure of the Universe} 
Princeton University Press

\bibitem[Rauch \etal\ (1997)]{rauch97} 
{Rauch, M., Miralda-Escude, J., Sargent, W. L. W., Barlow,
T. A., Weinberg, D. H., Hernquist, L., Katz, N., Cen, R., Ostriker,
J. P.} 1997, 
\textit{ApJ} 489, 7 

\bibitem[Schaye \etal\ (2000)]{schaye2000} 
{Schaye, J., Theuns, T., Rauch, M., Efstathiou, G., 
Sargent, W.L.W.} 2000, 
\textit{MNRAS} 318, 817



\bibitem[]{viel2002}
{Viel, M., Matarrese, S., Mo, H. J., Haehnelt, M., Theuns, T.} 2002,
\textit{MNRAS} 329, 848


\bibitem[]{viel2004}
{Viel, M.,  Haehnelt, M. G., Springel, V.} 
2004 \textit{MNRAS} 354, 684
\end{thebibliography}
\end{document}